\documentclass[aps,prl,showpacs,twocolumn]{revtex4}
\usepackage{mathrsfs}
\usepackage{graphicx}

\begin{document}

\title{On the quantization accuracy of the acoustoelectric current}
\author{Xiang-Bai Chen,$^{1,2,}$\footnote{Electronic address: 
xchen@ewha.ac.kr} 
Xiang-Song Chen,$^{1,}$\footnote{Electronic address: cxs@scu.edu.cn}
and Jie Gao$^1$} \affiliation{$^1$Department of
Physics, Sichuan University, Chengdu
610064, China\\
$^2$Department of Physics, Ewha Womans University, Seoul 120-750,
South Korea}
\date{\today}

\begin{abstract}
By deriving analytical formulae for the quantization accuracy of the
acoustoelectric current, we reveal that: 1) the flatness
of the current plateau for the typical present devices has reached
the theoretical limit of about 100ppm over a 1/1000 change of the
gate voltage; 2) increasing the transport channel length, and
counterintuitively, increasing the acoustic wavelength as well,
would improve the quantization accuracy, and very promisingly up to
0.01ppm as required for a quantum current standard. 
\end{abstract}

\maketitle

{\em Introduction}: The observation of a quantized electric current
driven by surface acoustic wave (SAW) in 1996 \cite{Shil96}
visualizes a possible quantum current standard. The device comprises
a split gate build on a GaAs-AlGaAs heterostructure containing a
two-dimensional electron gas (2DEG). A negative voltage $V_g$
applied to the split gate depletes the 2DEG beneath the gate and
leaves a one-dimensional (1D) transport channel. When pinched off,
the 1D channel becomes a saddle-like potential barrier. At large
wave amplitude, the SAW-induced potential minima become moving
quantum dots (QDs), which capture electrons within the dots and drag
them through the potential barrier. The important finding of Ref.
\cite{Shil96} is that such acoustoelectric current displays plateaux
over certain range of $V_g$ and/or SAW power, and the plateaux fit
well into the quantized expression $I=Nef_{\rm SAW}$, with $N$ an
integer, $e$ the electron charge, and $f_{\rm SAW}$ the SAW
frequency in the GHz band.

Qualitatively, this quantization phenomenon is explained by the
Coulomb-blockade effect: At a certain depth the QD can only hold a
fixed number of electrons, because excess occupation is prohibited
by significant Coulomb repulsion among electrons within the small
QD. The detailed electron transport process and possible error
mechanism in such devices have been studied theoretically but quite
controversially, in both classical dynamics \cite{Robi01} and
quantum mechanism \cite{Aizi98,Gumb99,Flen99,Maks00,Guo06}.
Sometimes, a very accurate current quantization (better than
$10^{-10}$) is predicted \cite{Maks00}. However, the flatness of the
current plateau achieved in the original \cite{Shil96} and
subsequent experiments
\cite{Taly97,Cunn99,Flet03,Gloo04,Robi05,Kata06} never goes beyond
$\sim$ 100ppm$/10^{-3}V_g$, far below the level of $\sim 10^{-8}$
required in metrology.

In this paper, we aim at a clear clue on what factors determine the
quantization accuracy of the acoustoelectric current. We start with
a novel analysis of the electron transport process and possible
error mechanisms by clarifying on the different mechanisms for
electrons to leave a QD. Then by some approximation techniques, we
derive analytical formulae for the quantization accuracy. These
formulae reveal clearly the limitation of the typical present
devices and the promising possibility of improving the accuracy for
metrological use.

{\em The electron transport process and error mechanisms}: Since the
quantization of acoustoelectric current is essentially the
quantization of electron number in each QD, we find it very
illuminating to first study how the electrons can escape a QD. We
clarify that there are three such escape mechanisms when the QD
climbs up the potential barrier:

\begin{enumerate}
\item[(i)] directly flowing out when the QD reduces its depth;
\item[(ii)] back tunneling through the rear wall of the QD;
\item[(iii)] non-adiabatic transition into excited states, followed
by an easier tunneling.
\end{enumerate}

Mechanism (i) occurs because the depth of the QD depends on the
slope of the potential barrier, which is not uniform and typically
has a maximum at the waist of the barrier. For mechanism (iii), the
excited state can in principle be a free state so that the electron
escapes the QD directly, but we focus on the dominant case of
transiting to a higher-energy bound state in the QD, and the
electron escapes via an enhanced tunneling probability.

With the above clarification, we divide the electron transport
process through the potential barrier into three segments,
characterized by the location of the QD:

\begin{enumerate}
\item[(a)] before the steepest ($\mathcal {S}$) point of the barrier;
\item[(b)] from the $\mathcal {S}$ point to the top of the barrier;
\item[(c)] after the top of the barrier.
\end{enumerate}

In segment (a), all three escape mechanisms are in effect. The QD
becomes shallowest at the $\mathcal {S}$ point, so in segment (b)
electrons can no longer flow out directly and only mechanisms (ii)
and (iii) remain. Segment (c) is irrelevant for us because all
escape mechanisms effectively kick the electrons into the drain
instead of the source.

Let us denote $N_{X}$ as the QD occupation number at the entrance of
the 1D channel, $N_{\mathcal{S}}$ as that at the $\mathcal {S}$
point, and $N_T$ as that at the top of the barrier. As we explained
above, $N_T$ can be take as the final number of transported
electrons per QD, so the acoustoelectric current is
\begin{equation}
I=N_T ef_{\rm SAW}. \label{I}
\end{equation}
Variation in $N_T$ means an error in the current quantization. It
should be noted that $N_T$ can only differ slightly from
$N_{\mathcal{S}}$, by the ``mild'' escape mechanisms (ii) and (iii).
In contrast, $N_{\mathcal{S}}$ may differ dramatically from $N_{X}$,
because in segment (a) the QD reduces its depth greatly, so that
electrons can escape in a ``wild'' manner of flowing out directly.
We write
\begin{equation}
N_T=(1-P_t-P_v) N_{\mathcal{S}}. \label{NT}
\end{equation}
Here $P_t$ is the (small) probability for an electron to escape the
QD in segment (b) by direct tunneling, and $P_v$ is the (small)
probability for an electron to first transit to excited states by
non-adiabatic correction, then escape the QD by an easier tunneling
due to higher energy. From Eq. (\ref{NT}), we can define two {\it
additive} error mechanisms in the quantization of $I=N_T ef_{\rm
SAW}$. First, the ``wild'' reduction from $N_X$ to $N_{\mathcal{S}}$
could be inaccurate, which results in an indefinite
$N_{\mathcal{S}}$. Second, even if $N_{\mathcal{S}}\equiv 1$, viz.,
exactly one electron remains in each QD at the $\mathcal {S}$ point,
this single electron may still escape the QD in segment (b) by
non-zero $P_t$ and $P_v$.

Investigation of the ``wild'' reduction from $N_X$ to
$N_{\mathcal{S}}$ necessarily involves many-electron dynamics. It
has been calculated numerically with classical dynamics in Ref.
\cite{Robi01}. And quantum mechanically, Ref. \cite{Flen99} has
performed a phenomenological study of this process at the very
entrance to the 1D channel. In the following, we concentrate on the
``mild'' reduction from $N_{\mathcal{S}}$ to $N_T$, starting with
$N_{\mathcal{S}}=1$, which is then just a one-particle problem.
Experimentally, the first current plateau (corresponding to one
electron transfer per SAW cycle) usually has the best quality. As we
will see below, the escape probability $(P_t+P_v)$ alone explains
the observed quantization accuracy very well. This means that the
error in the reduction from $N_X$ to $N_{\mathcal{S}}$ is at most of
the same order as $(P_t+P_v)$.

{\em Back tunneling} of electrons out of a moving QD has been
studied numerically in Refs. \cite{Aizi98,Gumb99,Guo06}. Here, we
develop some approximation methods which allow us to estimate this
effect analytically. Let us first look at the tunneling probability
given by the familiar WKB formula:
\begin{equation}
P=\exp{\left[-\frac{2}{\hbar}\int_{x_1}^{x_2} dx
\sqrt{2m^*(V(x)-E)}\right]}. \label{WKB}
\end{equation}
Here $m^*=0.067m_e$ is the effective electron mass, $V(x)$ describes
the QD rear wall to penetrate, $E$ is the electron energy, and
$x_{1,2}$ are the two transition points at which $V(x)=E$. It should
be noted that $P$ in Eq. (\ref{WKB}) is derived as a transmission
coefficient, viz., the intensity ratio of the out-going and incident
beams. In other words, $P$ is the penetration probability {\it per
collision}. If {\it an} electron captured in a QD collides with the
QD rear wall $\Gamma$ times per unit time, then the total
probability for it to tunnel out of the QD in a time duration $[t_1,
t_2]$ is
\begin{equation}
P_t =\int_{t_1}^{t_2} P\Gamma dt. \label{Pt0}
\end{equation}
The collision frequency $\Gamma$ is just the electron oscillation
frequency in the QD. If we approximate the electron motion in the QD
as a harmonic oscillator with angular frequency $\omega_e$, then
$\Gamma =f_e =\omega_e/2\pi$, which is typically much higher than
the SAW frequency $f_{\rm SAW}$. Evidently, Eq. (\ref{Pt0}) is
equally valid if $P$ and $\Gamma$ vary with time.

Eq. (\ref{WKB}) tells us that the back-tunneling probability $P$ is
largely determined by the height and width of effective part of the
QD rear wall lying above the electron energy. To examine these
parameters, we make a reasonable approximation by considering the
superposition of a sinusoidal potential wave $A\sin(2\pi x/\lambda)$
and a straight potential slope $\kappa x$, with $\kappa$
corresponding to the local slope of the real potential barrier. The
largest derivative of the sinusoidal wave, $2\pi A/\lambda$, must
exceeds $\kappa$, otherwise no QD can be formed. In a recent paper
\cite{Astl07}, the SAW-induced wave amplitude $A$ was estimated to
be 25 meV. In contrast, to make $N_{\mathcal{S}}=1$ the depth
$\Delta$ of the QD at the $\mathcal{S}$ point must be smaller than
the charging energy of the QD, which is estimated to be $\sim$ 1meV
\cite{Astl07,Flen99}. Since now $\Delta\ll A$, the slope $\kappa_0$
at the $\mathcal{S}$ point must be very close to $2\pi A/\lambda$ so
as to produce a superposed QD much shallower than $A$. We therefore
define
\begin{equation}
\frac{\kappa_0}{2\pi A/\lambda} \equiv 1-\eta, \label{eta}
\end{equation}
with $\eta$ a small parameter. Then, a slight algebra shows that the
depth $\Delta$ of the QD is approximated by
\begin{equation}
\Delta\simeq \frac{3}{\sqrt{2}}\sqrt{\eta^3} A, \label{Delta}
\end{equation}
and the effective width of the QD rear wall is approximated by
\begin{equation}
W\simeq\frac{3\sqrt{2}}{2\pi} \sqrt{\eta} \lambda. \label{W}
\end{equation}
By symmetry, $W$ is also the effective diameter of the QD. We see
that at the $\mathcal {S}$ point of the potential barrier, the size
of the QD is much smaller than the SAW wavelength.

The parameter $\eta$ is not free. It can be fixed as follows: To
permit only one electron in the QD, its depth $\Delta$ must not
exceed the charging energy $E_c\sim {e^2}/r\epsilon_{\rm eff}$, with
$r=W/2$ the effective radius of the QD and $\epsilon_{\rm eff}$ the
effective dielectric constant in the QD. From $\Delta \le E_c$, we
have
\begin{equation}
\eta\le \sqrt{\frac{ 4 \pi}{9A\lambda}\frac{e^2}{\epsilon_{\rm
eff}}}. \label{subtle}
\end{equation}
The upper limit corresponds to the best quantization accuracy. To
have a quantitative feeling of these parameters, take $A=25$meV,
$\lambda=1\mu$m, and tentatively $\epsilon_{\rm eff}=13$, we find
\begin{equation}
\eta \le 0.079, ~~~ \Delta \le 1.2 {\rm meV}, ~~~ W\le 0.19\mu{\rm
m}. \label{parameter}
\end{equation}
They are all consistent with our preassumptions.

There is a hidden subtlety in Eq. (\ref{subtle}) which calls for
special attention: the choice of $\epsilon_{\rm eff}$. Quite often,
it is argued that since the 2DEG resides very close to the surface,
(viz., with a distance $d\sim 0.1\mu$m much smaller than the SAW
wavelength) $\epsilon_{\rm eff}$ is roughly half of the value for a
body material. However, Eq. (\ref{parameter}) indicates that the
size of the QD is also much smaller than the SAW wavelength, and
actually comparable to $d$, thus within the QD $\epsilon_{\rm eff}$
can be close to the value ($\simeq 13$) for a body material.

To apply Eq. (\ref{WKB}), we still need the electron energy inside
the QD. We approximate the electron motion as a harmonic oscillator
with an angular frequency $\omega_e$:
\begin{equation}
m^*\omega_e^2=A(\frac{2\pi}{\lambda})^2\sqrt{1-(\frac{\kappa_0}{2\pi
A/\lambda})^2}\simeq A(\frac{2\pi}{\lambda})^2 \sqrt{2\eta}.
\label{omega}
\end{equation}
A quick order-of-magnitude estimation of the tunneling probability
can be obtained by approximating the QD rear wall as a triangle with
height $\Delta$ and width $W$:
\begin{equation}
P\sim \exp\left[-\frac 2 \hbar\cdot \frac 23 (\frac{\Delta
-E}{\Delta}W)\sqrt{2m^*(\Delta-E)}\right]. \label{Psim}
\end{equation}
Here $E\simeq (n+\frac 12) \hbar \omega_e$ is the electron energy
relative to the bottom of the QD, and the ground state has $n=0$.

The final ingredient needed to compute $P_t$ is an explicit
expression of the potential barrier. We take it to be
\begin{equation}
V(x)={V_0}/\cosh^2(x/L). \label{V}
\end{equation}
The effective width of this potential barrier is $2L$, the steepest
point is at $x_0\simeq -0.66L$, with a slope
\begin{equation}
\kappa_0\equiv C_1V_0/L\simeq 0.77V_0/L. \label{K0}
\end{equation}
The total probability for an electron to tunnel out in segment (b)
is then
\begin{equation}
P_t=\int_{x_0}^0 P(x) \frac{\omega_e}{2\pi} \frac{dx}{v_s},
\label{Pt}
\end{equation}
where $v_s\simeq 2800$m/s is the SAW speed.

There are three free parameters in the above expressions: the SAW
wavelength $\lambda$, the barrier width $2L$, and the barrier height
$V_0$. From Eqs. (\ref{eta}-\ref{K0}), the tunneling probability at
the $\mathcal{S}$ point can be expressed as
\begin{equation}
P\sim \exp\left[-C_t\lambda^{\frac 14}L^{\frac 18}V_0^{-\frac
18}\left(1-C_e\lambda^{-\frac 14}L^{-\frac 18}V_0^{\frac
18}\right)^{\frac 32}\right]. \label{master1}
\end{equation}
This is our first master equation. Here $C_t$ and $C_e$ are
straightforwardly computed coefficients independent of $\lambda$,
$L$, and $V_0$. We postpone discussing the implications of Eq.
(\ref{master1}), and turn to the other error mechanism:

{\em Non-adiabatic excitation}: This effect can be most conveniently
studied by adopting a coordinate system co-moving with the SAW.
Because the potential barrier is not uniform, even in this co-moving
coordinate system the electron feels a varying external field, which
can lead to quantum transitions. Namely, an electron in state $a$ at
time $t=0$ has a non-zero probability $P_{ba}(t)$ to be in another
state $b$ at a later time $t$. To first order, $P_{ba}(t)$ is given
by \cite{Bran89}
\begin{equation}
P_{ba}(t)\simeq 4 \hbar ^{-2} \omega_{ba}^{-4}
\left|\left\langle\frac{\partial H}{\partial t} \right\rangle_{ba}
\right|^2 \sin^2[\omega_{ba}t/2]. \label{Pba}
\end{equation}
Here $\omega_{ba}(t) =[E_b(t)-E_a(t)]/\hbar$, with $E_{a,b}(t)$
eigenvalues of the instantaneous Hamiltonian $H(t)$.
$\left\langle\frac{\partial H}{\partial t} \right\rangle_{ba}$ is
the matrix element of $\frac{\partial H}{\partial t}$ between the
states $b$ and $a$. Because the electron oscillation frequency is
much higher than the SAW frequency, the factor
$\sin^2[\omega_{ba}t/2]$ in Eq. (\ref{Pba}) oscillates very quickly.
By approximating the electron motion in the QD as a harmonic
oscillator, our problem resembles the textbook example of a charged
harmonic oscillator in a time-dependent electric field
\cite{Bran89}, for which only the transition into the first excited
state is relevant, and the transition probability $P_{10}$ can be
expressed as
\begin{equation}
P_{10}\simeq \left(\frac{v_s}{v_e}\right)^2
\left(\frac{1}{m^*\omega_e^2}\frac{d^2V}{dx^2}\right)^2
\sin^2[\omega_e t/2]. \label{Pv}
\end{equation}
Here $v_e\equiv \sqrt{2\hbar\omega_e/m^*}$ is an effective electron
speed in the QD, with $\omega_e$ from Eq. (\ref{omega}). To see the
dependence of $P_{10}$ on the experimental parameters, we write
\begin{equation}
d^2V/dx^2\equiv C_2 V_0/L^2,
\end{equation} with the coefficient $C_2$ independent of $V_0$ and $L$.
Following the same steps as in deriving Eq. (\ref{master1}),
$P_{10}$ can be expressed as
\begin{equation}
P_{10}\simeq C_v\lambda^{\frac{15}{4}}L^{-\frac{17}{8}}V_0^{\frac 18
}\sin^2[\omega_e t/2]. \label{master2}
\end{equation}
This is our second master equation. Here $C_v$ is another
coefficient independent of $\lambda$, $L$, and $V_0$.

{\em Quantization accuracy for the present devices}: We now plug in
explicit parameters to estimate the theoretical limit of the
quantization accuracy of the acoustoelectric current. In a typical
experiment \cite{Shil96,Taly97,Cunn99,Flet03,Gloo04,Robi05,Kata06},
one has $\lambda=2L=1\mu$m. There is just one parameter to tune: the
gate voltage (which determines the barrier height $V_0$) or the SAW
amplitude. But this parameter does not bring much freedom either.
$V_0$ must be at least several tens of meV in order to be well above
the electron Fermi energy. Furthermore, our two master equations
(\ref{master1}) and (\ref{master2}) show a rather week dependence on
$V_0$. For the experimentally estimated $A=25$meV \cite{Astl07},
Eqs. (\ref{eta}) and (\ref{K0}) give $V_0\simeq 94$meV, in
consistent with our expectation. Then, Eq. (\ref{Psim}) or
(\ref{master1}) gives the tunneling probability of the ground-state
electron at the $\mathcal {S}$ point: $P_0\sim 10^{-3}$, and Eq.
(\ref{Pt}) gives the total probability for this electron to tunnel
out in segment (b): $P_t \simeq 1.3P_0$. This integrated tunneling
probability is of the same order as the tunneling probability at the
$\mathcal {S}$ point because $P$ drops very quickly as $x$ moves
away from the $\mathcal {S}$ point.

The situation for non-adiabatic transition requires some caution.
For an electron to eventually escape the QD by this mechanism, the
transition must be followed by an easy tunneling. The tunneling
probability is largest at the $\mathcal {S}$ point, but Eq.
(\ref{Pv}) says that the transition probability is related to the
second derivative $dV^2/dx^2$, which is zero at the $\mathcal {S}$
point. At first thought this seems to imply that the non-adiabatic
mechanism is not relevant. However, we note that our whole
discussion for non-adiabatic transition can as well be applied to
the center-of-mass motion of the interacting electrons in segment
(a), where $d^2V/dx^2$ can be significant. This means that during
this period the center-of-mass motion of electrons may transit to an
excited state, by a probability of the same order as given by Eq.
(\ref{Pv}). The excited electrons will then escape the QD by the
significant tunneling probability near the $\mathcal {S}$ point,
leading to an error in current quantization. Therefore, to safely
control this kind of quantization error, we should estimate $P_v$ as
the product of the maximum $P_{10}$ in segment (a) with $P_1$ at the
$\mathcal {S}$ point, where $P_1$ is the tunneling probability of
the electron in the first excited state. A slight calculation with
Eq. (\ref{Pv}) gives $P_{10}\sim 2\times 10^{-4}$, and Eq.
(\ref{Psim}) gives $P_1\sim 0.5$. Note that the WKB formulae is only
valid at very small tunneling probability, therefore a $P_1\sim 0.5$
should be effectively taken as one (viz., 100\% electrons escape).
So we finally estimate $P_v$ in Eq. (\ref{NT}) to be $\sim 2\times
10^{-4}$.

Putting $P_t$ and $P_v$ together, we see that the acoustoelectric
current at the first plateau deviates from the quantized value
$I_1=ef_{\rm SAW}$ by at least $1.5\times 10^{-3}$. This absolute
deviation is hard to detect due to the zero-point drift in measuring
the electric current. What is easier to detect is the slope of the
current plateau. By fixing $A=25$meV and varying $V_0$, the slope is
computed to be $dI/dV_0\sim 10^{-4}I_1/10^{-3}V_0$. Since $V_0$ is
proportional to the gate voltage $V_g$, this means that the
acoustoelectric current varies by 100ppm as $V_g$ varies by 1/1000.
This agrees very well with the best experimental result
\cite{Cunn99}.

{\em Tuning the parameters for better accuracy}: From our two master
equations (\ref{master1}) and (\ref{master2}), we see that the
quantization accuracy of acoustoelectric current is essentially set
by the device parameters $\lambda$ and $L$. A longer transport
channel, viz, a wider potential barrier, always reduces the error in
current quantization. And quite counterintuitively, a longer SAW
wavelength may also be advantageous, because as $\lambda$ increases
the tunneling probability falls exponentially while the
non-adiabatic transition probability increases in a power law.

A larger $\lambda$ also brings a hidden benefit that the QD size may
indeed considerably exceed $d$, the distance of the 2DEG to the
surface, thus the effective dielectric constant reduces and the
Coulomb-blockade effect ehances. E.g., if we take $\lambda=L=5\mu$m,
$V_0=90$meV, and pre-assume that $r=\frac 12 W\ll d$ so that
$\epsilon_{\rm eff}\simeq 13/2$, we get $r\simeq 0.46\mu$m, in
consistent with the preassumption. Proceeding with $\epsilon_{\rm
eff}\simeq 13/2$, Eqs. (\ref{Psim}) and (\ref{Pt}) give $P_0\sim
P_t\sim 10^{-14}$ and $P_1\sim 10^{-10}$, Eq. (\ref{Pv}) gives
$P_{10}\sim 10^{-3}$, and thus $P_v\sim 10^{-13}$. The slope of the
current plateau is computed to be $dI/dV\sim 10^{-13}
I_1/10^{-3}V_0$. These accuracies are far enough for metrological
use, which requires 0.01ppm.

{\em Summary}: We performed a novel examination of the quantization
accuracy of the acoustoelectric current. By clarifying the electron
transport process and error mechanisms, we obtain a clear clue on
the theoretical limit of the quantization accuracy, expressed by
several analytical approximate formulae. We have shown unambiguously
that the so far achieved quantization accuracy is already the
theoretical limit of the present devices with $\lambda=2L\sim
1\mu$m. To improve the accuracy to meet metrological requirement, it
is necessary, and very promising, to have a larger
 $L$ and/or $\lambda$, e.g., of $5\mu$m.

This work is supported by the China NSF under Grant No. 60436010 and
the State-Support Program of Science and Technology under Grant No.
2006BAF06B09.

\end{document}